\documentclass{IEEEtran}
\usepackage{graphicx} 
\usepackage{color}
\usepackage{comment}
\usepackage{soul}
\usepackage{amsmath}
\usepackage{amssymb}
\usepackage{subcaption}
\usepackage{caption}
\usepackage{subfiles}
\usepackage{cite}
\usepackage{url}

\newcommand{\blue}[1]{\textcolor[rgb]{0,0,0}{#1}}

\title{Modelling and Performance Analysis of Non-Primary Channel Access in Wi-Fi Networks}

\author{
\IEEEauthorblockN{Boris Bellalta$^\flat$, Francesc Wilhelmi$^\flat$, Lorenzo Galati Giordano$^\star$, and Giovanni Geraci$^\sharp$$^\flat$\\ \vspace{0.2cm}}

\IEEEauthorblockA{$^\flat$\emph{Department of Engineering, Universitat Pompeu Fabra, Barcelona, Spain}}\\
\IEEEauthorblockA{$^\star$\emph{Radio Systems Research, Nokia Bell Labs, Stuttgart, Germany}}\\
\IEEEauthorblockA{$^\sharp$\emph{Nokia Standards, Spain}}
}

\date{}

\begin{document}

\maketitle

\begin{abstract}
\blue{This paper aims to improve our understanding of the performance of the Non-Primary Channel Access (NPCA) mechanism, a new feature introduced in IEEE 802.11bn to enhance spectrum utilization in Wi-Fi networks.} NPCA enables devices to contend for and transmit on the secondary channel when the primary channel is occupied by transmissions from an Overlapping Basic Service Set (OBSS). We develop a Continuous-Time Markov Chain (CTMC) model that captures the interactions among OBSSs in dense Wireless Local Area Network (WLAN) environments when NPCA is enabled, incorporating new NPCA-specific states and transitions. In addition to the analytical insights offered by the model, we conduct numerical evaluations and simulations to quantify NPCA’s impact on throughput and channel access delay across various scenarios. Our results show that NPCA can significantly improve throughput and reduce access delays in favorable conditions for BSSs that support the mechanism. Moreover, NPCA helps mitigate the OBSS performance anomaly, where low-rate OBSS transmissions degrade network performance for all nearby devices. However, we also observe trade-offs: NPCA may increase contention on secondary channels, potentially reducing transmission opportunities for BSSs operating there. \blue{Overall, the proposed modeling approach offers a foundation for analyzing, optimizing, and guiding the development of NPCA in next-generation Wi-Fi networks.}
\vspace{2mm}
\end{abstract}

\begin{IEEEkeywords}
Non-primary channel access, NPCA, IEEE 802.11bn, WLAN, Wi-Fi 8. 
\end{IEEEkeywords}


\section{Introduction}
\label{sec:Intro}

\noindent \textit{The context}: Wi-Fi channel widths have increased over time, from the initial 20 MHz to 320 MHz, going through 40, 80, and 160 MHz. The adoption of wider channels aims to enhance network performance by increasing transmission rates, improving throughput, and reducing latency~\cite{geraci2025wi}. However, as a result of current channel sensing rules based on Listen Before Talk (LBT), this comes with the drawback of higher contention with neighboring networks operating on the same channel or overlapping portions of it~\cite{barrachina2021wi}.

Channels wider than 20 MHz consist of a mandatory 20 MHz primary channel---where all devices within the same Basic Service Set (BSS) perform channel access contention (e.g., backoff countdown)---and one or more secondary channels. Secondary channels can only be used if they are idle at the time a transmission opportunity is secured via the primary channel. If one or more secondary channels are occupied by an Overlapping BSS (OBSS), they are effectively bypassed by puncturing them thanks to the preamble puncturing feature introduced in IEEE 802.11ax (Wi-Fi 6)~\cite{khorov2018tutorial}. While this reduces the effective channel width, it still allows devices to transmit using the instantaneous available spectrum. 

\begin{figure}
    \centering
    \includegraphics[width=0.95\linewidth]{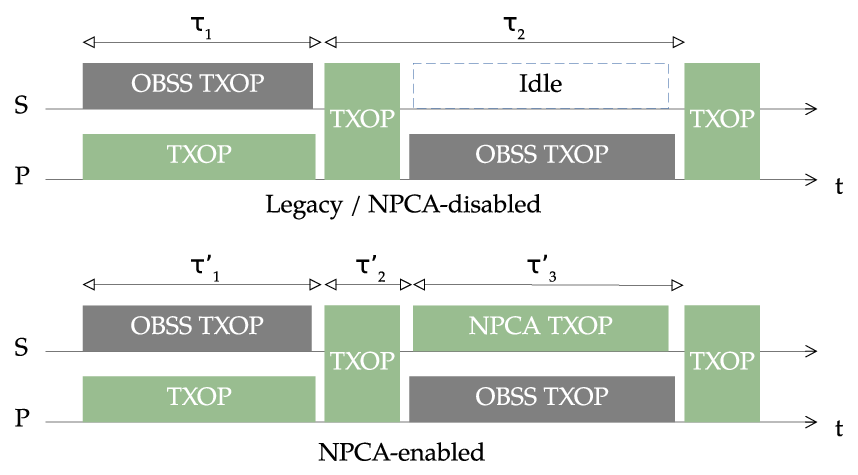}
    \caption{Legacy (top) vs. NPCA (bottom) operation, where `P' and `S' denote the primary and secondary channel, respectively, e.g., an 80 MHz channel each. The figure illustrates how NPCA enables near-continuous transmissions, resulting in higher throughput and reduced channel access delays, as reflected by the lower times between consecutive channel accesses ($\tau$ values).}
    \label{fig:NPCA_Motivation}
\end{figure}

\vspace{0.25cm}
\noindent \textit{The problem}: Wi-Fi currently lacks a solution for scenarios where the primary channel of a BSS is occupied by an OBSS transmission. In such cases, even if all the secondary channels are available, devices in the affected BSS must defer their transmissions until the primary channel becomes idle again. This results in reduced channel access opportunities and inefficient spectrum utilization, as illustrated in Fig.~\ref{fig:NPCA_Motivation} (upper part). 

\vspace{0.25cm}
\noindent \textit{The 802.11bn solution}: \blue{To address this issue, IEEE 802.11bn~\cite{galati2024will,valwi} is developing a new feature called Non-Primary Channel Access (NPCA).} The concept is straightforward: when a BSS detects an OBSS transmission on its primary channel, NPCA-compliant devices temporarily switch to a secondary channel, designated as the NPCA primary channel, to perform channel contention and initiate transmissions while the primary is busy. This enables the BSS to utilize idle secondary channels effectively. The benefits of this approach are illustrated in Fig.~\ref{fig:NPCA_Motivation} (lower part). With NPCA, the target BSS increases its chances of accessing the channel, which is crucial for latency-sensitive traffic, while ensuring the full utilization of available spectrum.

\vspace{0.25cm}
\noindent \textit{Our contribution}: In this paper, we investigate the potential throughput and channel access delay gains of NPCA. To achieve this, we use Continuous Time Markov Chains (CTMCs) to model the system behavior with and without NPCA. CTMCs have been widely and successfully applied to analyze the performance of complex Carrier Sense Multiple Access with Collision Avoidance (CSMA/CA) networks~\cite{boorstyn1987throughput, liew2010back, laufer2015capacity}, including scenarios involving advanced IEEE 802.11 features such as channel bonding~\cite{bellalta2015interactions, faridi2016analysis, barrachina2019dynamic} and spatial reuse~\cite{wilhelmi2021spatial, wilhelmi2023throughput}. Specifically, we resort to CTMC models to capture NPCA operation, leveraging their suitability for studying CSMA/CA-based networks in general and Wi-Fi in particular.

The main contributions of this paper are as follows:
\begin{enumerate}
    \item \blue{We describe the principles of NPCA operation based on IEEE 802.11bn discussions, and propose to support multi-NPCA consecutive transmissions, aiming to better exploit large NPCA opportunities.}  
    \item We extend CTMC 802.11 models to incorporate NPCA operation, enabling detailed performance analysis. \blue{As part of our modeling approach, we examine the key design considerations and underlying assumptions, detailing their implications and how they influence the achievable results and their significance, thereby laying the groundwork for future modeling efforts.}
    \item We use the CTMC model to evaluate NPCA and demonstrate that it effectively provides throughput and channel access gains for the NPCA-enabled BSS while remaining completely transparent to OBSS networks operating on the NPCA-enabled BSS primary channel.  
    \item We uncover the relationship between the Transmission Opportunity (TXOP) duration and NPCA performance, highlighting a trade-off between NPCA throughput gains and the Aggregated MAC Protocol Data Unit (A-MPDU) size.  
    \item We show that NPCA mitigates the OBSS performance anomaly, particularly when OBSS transmissions are disproportionately long due to using low Modulation and Coding Scheme (MCS), while the target BSS employs high MCSs. In such cases, throughput gains exceeding $\times2$ can be achieved.  
    \item We illustrate that, in the presence of OBSSs operating on the secondary channel of the target NPCA BSS, enabling NPCA has a significant impact on the throughput distribution between the target BSS and the OBSSs.  
\end{enumerate}

\section{Non Primary Channel Access} \label{Sec:NPCA}

\begin{figure*}[th!]
    \centering
    \includegraphics[width=0.80\linewidth]{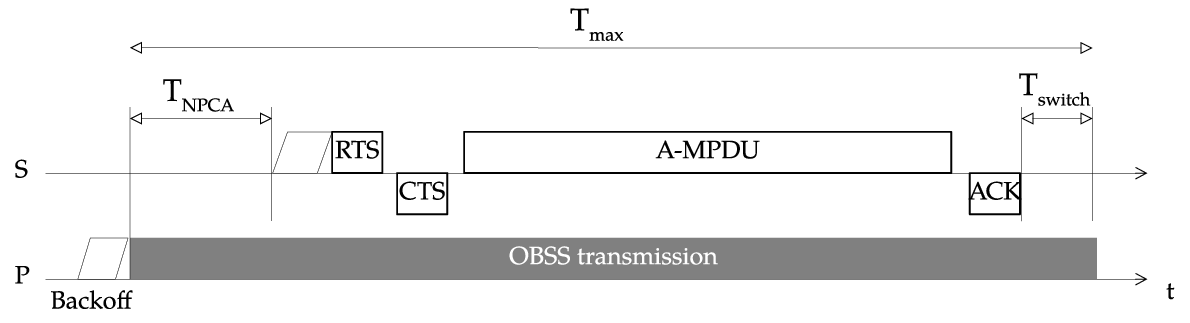}
    \caption{Non Primary Channel Access operation.}
    \label{fig:NPCA}
\end{figure*}

The operation of NPCA is illustrated in Fig.~\ref{fig:NPCA}, where the activity on primary and secondary channels of a given BSS is shown. In the context of NPCA, we refer to the secondary channel as the \textit{NPCA channel}. Additionally, the 20~MHz channel within the NPCA channel used for contention is referred to as the \textit{NPCA primary channel}.

When an OBSS transmission is detected on an NPCA-enabled BSS's primary channel, after receiving both Request-To-Send (RTS) and Clear-To-Send (CTS) frames, all NPCA-capable devices switch their control channel to a predefined 20 MHz secondary channel (i.e., the NPCA primary channel). The contention in the NPCA primary channel begins after a delay of $T_{\rm NPCA}$ from the start of the OBSS transmission. This delay $T_{\rm NPCA}$ corresponds to the duration of the RTS/CTS exchange. Once the backoff counter reaches zero, a transmission on the secondary channel is initiated following the default 802.11 rules.


A key requirement of NPCA is that transmissions must complete before the OBSS transmission ends, so that the BSS can resume operation on its default primary channel. This constraint is important because not all devices within the BSS are assumed to support NPCA. We also account for a switching delay, denoted $T_{\rm switch}$, which represents the time required to return from the NPCA primary channel to the default primary channel. 
As a result, the time available for an NPCA transmission opportunity is limited by the duration of the OBSS transmission, reduced by the sum of $T_{\rm NPCA}$ and $T_{\rm switch}$. These additional switching overheads reduce the effective time available for data transmission in NPCA, leading to slightly lower efficiency compared to legacy (non-NPCA) transmissions.

Finally, we make the following two assumptions regarding NPCA operation, both of which leverage the fact that the NPCA BSS is aware of the expected end time of the OBSS transmission occupying its primary channel:
\begin{enumerate}
    \item After switching to the NPCA primary channel due to detecting an OBSS transmission on the primary channel, if the NPCA primary channel is initially busy, the target BSS will continue to monitor it and attempt to transmit if it becomes idle before the NPCA opportunity ends.
    \item If a transmission on the NPCA channel concludes and there is still sufficient time to initiate another transmission, the target BSS will re-contend for access on it by executing a new backoff procedure on the NPCA primary channel. If successful, it will initiate another transmission, with its duration adjusted to fit within the remaining time of the OBSS transmission. 
\end{enumerate}
In both cases, the target BSS returns to its default primary channel before the OBSS transmission completes.

\section{System Model}
\label{Sec:Scenarios}

\begin{figure}[t!]
    \centering
    \includegraphics[width=0.95\linewidth]{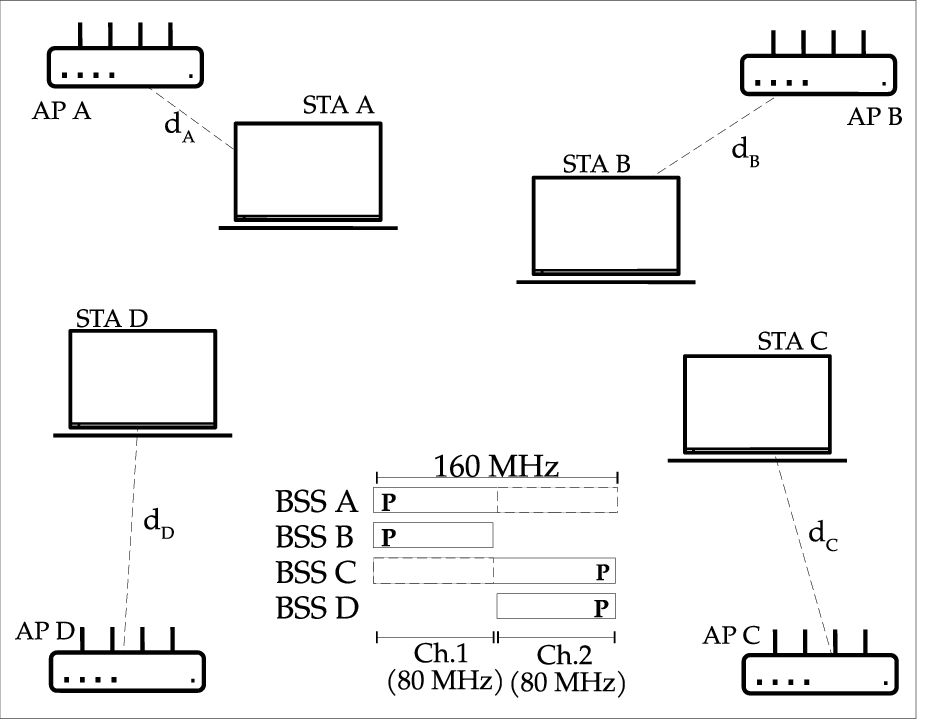}
    \caption{Four overlapping BSSs.}
    \label{fig:Scenario4BSSs_v2}
\end{figure}

To analyze the NPCA performance, we consider the Wi-Fi deployment depicted in Fig.~\ref{fig:Scenario4BSSs_v2}. The scenario consists of four BSSs, each consisting of one Access Point (AP) and one \blue{station (STA)}. The channel allocation for each BSS is detailed in the figure: BSS A and C utilize 160 MHz channels, while BSS B and D operate on 80 MHz channels. All devices, including APs and STAs, are within each other’s coverage area, eliminating the possibility of hidden terminals. For simplicity, the lower 80 MHz of the 160 MHz channel is referred to as channel~1 (Ch\#1), and the upper 80 MHz as channel~2 (Ch\#2). When APs A and C use the 160 MHz channel, it is indicated as utilizing both Ch\#1 and Ch\#2.

STAs are deployed at a distance $d$ from their respective APs. The TMB path loss model for the 5 GHz band in indoor office environments is considered~\cite{adame2019tmb}. The selected MCS values depend on the received power and range from MCS~11~(1024-QAM; 5/6) to MCS~1~(BPSK; 1/2). We use different transmission power values when transmitting over 80 MHz (20 dBm) and 160 MHz (23 dBm) channels to ensure that the same MCS is used regardless of the channel width. The maximum TXOP duration is set to $T_{\max}=5$ ms, encompassing the RTS/CTS exchange, data transmission, Block Acknowledgment (BACK), and inter-frame spaces. A Packet Error Rate (PER) of $0.1$ is assumed for all MCSs. All transmissions utilize two spatial streams for the data part, while one spatial stream is used for all control frames.

Only downlink traffic is considered, meaning that only APs transmit data packets. Throughout the paper, we use the terms AP and BSS interchangeably to refer to the transmitting device. Each data packet has a size of $L$ bits. A-MPDU packet aggregation is employed, with $N$ packets aggregated per transmission. The value of $N$ depends on the MCS, channel width, the maximum TXOP duration $T_{\max}$, and the maximum A-MPDU size $\Delta$. Specifically, $N \leq \min(M, \Delta)$, where $M$ is the maximum number of packets that can be transmitted within $T_{\max}$ for a given MCS and channel width. For example, using MCS 11, BSS A can transmit up to 968 data packets of 1400 bytes each (approximately 10 Mbits) within $T_{\max}$ on a 160~MHz channel. In contrast, with MCS 1 on an 80~MHz channel, BSS A can transmit up to 29 packets (approximately 325 kbits) during the same duration.

\blue{Regarding NPCA operation, we assume that a backoff instance is generated for each transmission, regardless of whether the BSS operates in legacy or NPCA mode, following the default IEEE 802.11 rules. In legacy mode, the backoff counter is paused whenever the primary channel is found to be busy, whereas in NPCA mode, it is paused only when both the primary and secondary channels are busy. Note that a BSS may switch between legacy and NPCA modes multiple times during a single backoff instance, until channel access is achieved.}

We study the performance of NPCA in three representative scenarios that are extracted from the deployment of Fig.~\ref{fig:Scenario4BSSs_v2}. These three scenarios cover the most important aspects needed to study NPCA performance. Considering more BSSs will only increase contention, proportionally scaling the performance each BSS can achieve, without providing further insights. Results in other scenarios can be easily extrapolated from the ones presented in this paper. 
\begin{enumerate}
    \item \textbf{Scenario I}: BSSs A and B are active. This scenario provides the most favorable conditions to evaluate NPCA throughput gains for BSS A, as its 80 MHz secondary channel remains always available. NPCA is expected to increase BSS A's channel access rate, enhancing throughput and reducing access delay.
    \item \textbf{Scenario II}: BSSs A, B, and D are active. Compared to Scenario I, activating BSS D occupies BSS A's secondary 80 MHz channel, limiting NPCA transmission opportunities. This scenario examines how activity on BSS A's secondary channel impacts NPCA gains.
    \item \textbf{Scenario III}: All four BSSs are active. This creates a symmetric scenario where BSS A and BSS C utilize NPCA to access channels 2 and 1, respectively. We investigate whether NPCA still provides performance gains in such a balanced setting.
\end{enumerate}

In Table~\ref{tab:peva_parameters}, we present the parameters used in the performance evaluation. Unspecified parameters follow the IEEE 802.11 specifications. If different values are considered during the evaluation, they will be indicated accordingly. For each data point in the results, 500 random instances of a given scenario have been simulated, with each instance lasting 10~s.

\begin{table}[t!]
    \centering
    \begin{tabular}{cc|cc}
      Parameter  & Value &  Parameter  & Value \\
      \hline
      CW$_{\min}$ & 16 & L & 1400 Bytes \\
      d & $\mathcal{U}[1,17]$ meters & MCSs (11ax) & [1-11] \\
      $P_{\rm tx}$ (80 MHz) & 20 dBm & $P_{\rm tx}$ (160 MHz) & 23 dBm \\
      $T_{\max}$ 	& 5 ms & Num SS. & 2 \\
      $\Delta$ & [1-1024] & PER & 0.1 \\
      $T_{\rm NPCA}$ & $0.136$ ms & $T_{\rm switch}$ & $16~\mu$s \\
      \hline
      OFDM symbol & 13.6~$\mu_s$ & Backoff slot & 9~$\mu$s \\
      DIFS & 34~$\mu$s & SIFS & 16~$\mu$s \\
      Leg. PHY pream. & 20$~\mu$s & PHY pream. & 100$~\mu$s \\
      RTS & 160 bits & CTS & 112 bits \\
      MAC header & 240 bits & BACK & 240 bits \\
      MPDU Del. & 32 bits & Tail Bits & 18 bits \\
      \hline
    \end{tabular}
    \caption{Value of the parameters used in the performance evaluation.}
    \label{tab:peva_parameters}
\end{table}

\section{Modeling NPCA with CTMCs}\label{sec:model}

CTMC models are widely adopted for their ability to effectively capture the complex, asynchronous interactions among devices sharing spectrum resources through CSMA/CA, as in Wi-Fi networks. \blue{These models have been validated against simulations in~\cite{liew2010back, bellalta2015interactions, bellalta2016throughput, faridi2016analysis, michaloliakos2016performance,barrachina2019komondor, barrachina2019dynamic, stojanova2021markov, wilhelmi2021spatial}, demonstrating their accuracy, representativeness, and consistency.}

A CTMC model captures the system's dynamics, enabling the analysis of its steady-state performance~\cite{boorstyn1987throughput}. In applying CTMCs to characterize Wi-Fi, we consider that a state $s \in \mathcal{\Omega}$---where $\mathcal{\Omega}$ is the set of all CTMC states---is defined as the set of active BSSs (i.e., BSSs that are concurrently transmitting), and we assume that channel access contention and transmission durations are governed by stochastic processes that follow exponential distributions.

\blue{However, CTMCs cannot accurately model collisions between contending devices. Specifically, because backoff durations are assumed to be exponentially distributed, the probability that two devices complete their backoff at exactly the same time is zero. As a result, simultaneous channel access---and hence collisions---are effectively excluded from the model. In our scenario, the impact of neglecting collisions is minimal, as the number of contenders is small---resulting in a low collision probability by default~\cite{bianchi2000performance}---and the use of RTS/CTS further minimizes collision duration, which is approximately 30 times shorter than that of a successful transmission. Consequently, neglecting collisions does not significantly affect the overall system performance (see Section~\ref{Sec:validation}).} Nonetheless, the impact of collisions can be incorporated in CTMCs-based models using the approach outlined in~\cite{bellalta2016throughput}.

The stationary distribution ($\vec{\pi}$) of a CTMC is derived by solving $\vec{\pi} \mathbf{Q} = 0$, where $\mathbf{Q}$ is the infinitesimal generator matrix of the stochastic process. Each element of $\mathbf{Q}$, denoted $Q_{i,j}$, represents the transition rate from state $i$ to state $j$. Forward transitions (e.g., initiating a transmission) occur at rate $\lambda$, while backward transitions (e.g., completing a transmission) occur at rate $\mu$.

\subsection{CTMC States}

To model our described scenario, which includes NPCA transmissions, we adopt the approach used in~\cite{faridi2016analysis,barrachina2019dynamic} for analyzing Dynamic Channel Bonding (DCB). However, in this work, we introduce a new type of state, referred to as an \textit{NPCA state}, where NPCA transmissions occur. A key characteristic of NPCA states is that, once the OBSS transmission ends, the system transitions to the same subsequent state that would follow the OBSS transmission in the absence of NPCA.

NPCA states become feasible when a BSS supporting NPCA detects that its primary channel is occupied by an OBSS transmission. If the secondary channel is idle, the NPCA-enabled BSS initiates (after contending) a transmission on that secondary channel, following the NPCA operation described in Section~\ref{Sec:NPCA}. The only difference in our model is that contention on the secondary channel is assumed to begin immediately after detecting the primary channel is busy, rather than after waiting for $T_{\mathrm{NPCA}}$ seconds. \blue{In such a situation, since the channel access probabilities of both the NPCA BSS and the OBSSs depend solely on their backoff parameters, this may result in higher throughput for the NPCA BSS at the expense of reduced throughput for the OBSSs, particularly if all of them use the same backoff configuration.}

\begin{figure*}
    \centering
    \includegraphics[width=0.95\linewidth]{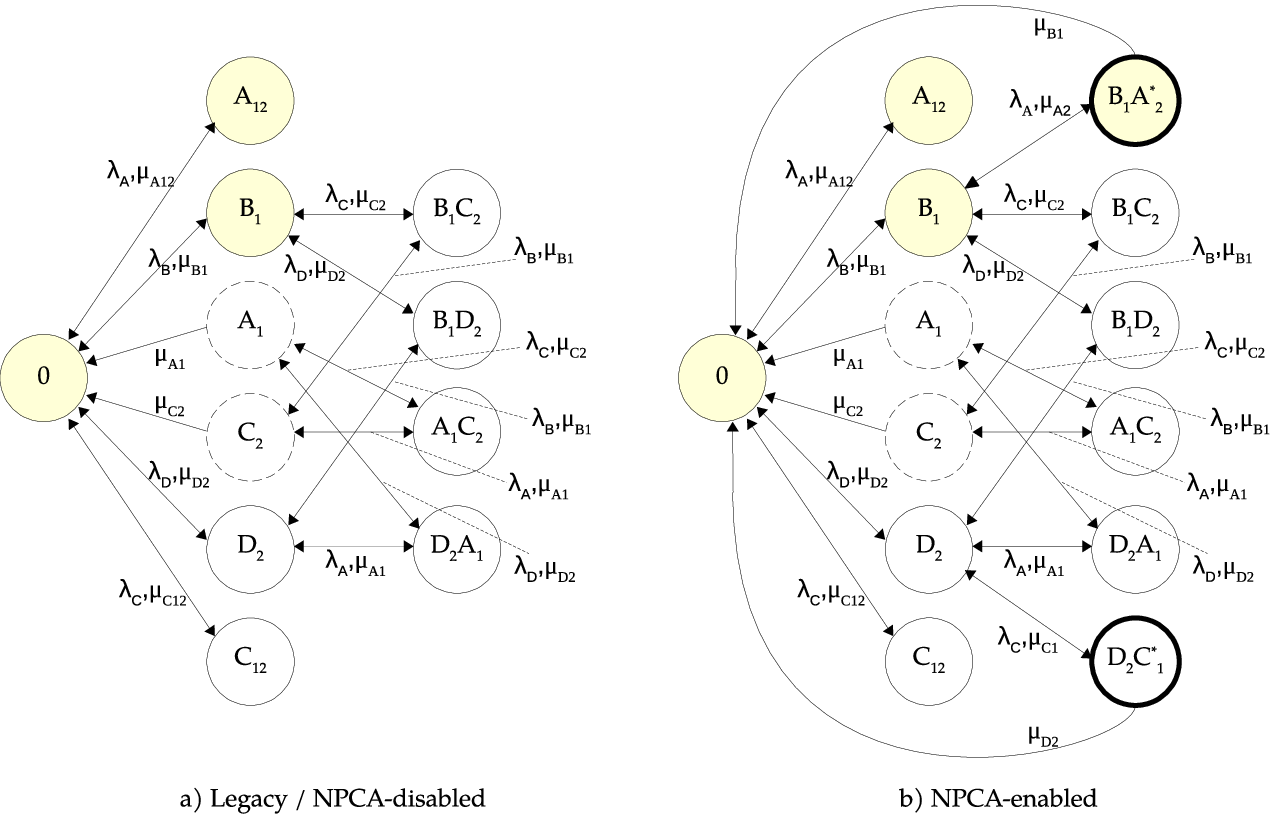}
    \caption{CTMC modeling the Wi-Fi deployment shown in Fig.~\ref{fig:Scenario4BSSs_v2} when a) Legacy (NPCA disabled), and b) NPCA is enabled.}
    \label{fig:CTMCsv3}
\end{figure*}

Fig.~\ref{fig:CTMCsv3} illustrates the CTMC for the Wi-Fi deployment depicted in Fig.~\ref{fig:Scenario4BSSs_v2}, comparing two scenarios: (a) when NPCA is not supported (or disabled), and (b) when NPCA is enabled. 
Each state is identified by the BSSs transmitting simultaneously (the capital letters), along with the specific channels each one uses (indicated by subscripts). For example, state $A_{12}$ denotes BSS A transmitting alone over Ch\#1 and~2 (a 160~MHz transmission), while state $B_1D_2$ indicates that BSS B and BSS D are transmitting concurrently---BSS B using Ch\#1 and BSS D using Ch\#2, each performing 80~MHz transmissions. States outlined with dashed lines correspond to Dynamic Channel Bonding (DCB) operations. As detailed in~\cite{faridi2016analysis,barrachina2019dynamic}, these states arise when a BSS transmits over only a portion of its allocated channel. DCB states cannot be reached from the idle state (state~$0$), since a BSS will not voluntarily access just a fraction of its allocated bandwidth when the entire channel is available. States in which only BSSs A and B are active are highlighted in yellow, corresponding to the active states of Scenario~I described in Section~\ref{Sec:Scenarios}.
Without NPCA~(Fig.~\ref{fig:CTMCsv3}.a), when BSS B (or BSS D) accesses Ch\#1 (or Ch\#2), the other channel can only be used by BSSs~C and D (or BSSs A and B, respectively). For instance, from state $B_1$, the system can only transition to states $B_1C_2$ or $B_1D_2$. When NPCA is enabled (Fig.~\ref{fig:CTMCsv3}.b), two additional states become available: $B_1A^*_2$ and $D_2C^*_1$. These states represent scenarios where BSS~A or BSS~C initiates a transmission on their NPCA channel while their respective primary channel is occupied by BSS~B or BSS~D. Transitions from these NPCA states return to state~$0$ upon the completion of the OBSS transmission that initially blocked access on the primary channel. As explained in Section~\ref{Sec:NPCA}, during this interval, multiple consecutive NPCA transmissions may occur if the NPCA-enabled BSS (e.g., BSS~A) completes its transmission quickly but still has packets queued in its buffer.

The CTMC states reveal that enabling NPCA increases contention, as illustrated in Fig.~\ref{fig:CTMCsv3}b with the higher number of forward transitions from states $B_1$ and $D_2$ compared to the same states in Fig.~\ref{fig:CTMCsv3}a, which entail that a bigger set of contenders is competing for the channel. The additional contention introduced by NPCA can be mitigated by configuring NPCA transmissions to use more conservative backoff parameters, such as a larger contention window~(CW). However, this question lies beyond the scope of this paper.

\subsection{CTMC Transitions}

\subsubsection{Channel access rate, $\lambda$}

The transmission attempt rate, i.e., how aggressively a device contends to access the channel, is defined as the reciprocal of the expected backoff time, given by
\begin{equation}
    \lambda = \frac{1}{E[\text{backoff}]} = \frac{2}{(\text{CW}-1)T_{\textrm{e}}} ,
    \label{eq:lambda}
\end{equation}
where $T_{\textrm{e}}$ is the duration of an empty slot (9 $\mu$s). 

\subsubsection{Transmission rate, $\mu$}

The mean transmission duration $\mathbb{E}[T_s] = 1/\mu$ represents the time a device occupies the channel after gaining access. We follow the frames and intervals specified by the IEEE 802.11 protocol. In particular, the duration of an A-MPDU transmission comprising $N$ packets is given by:
\begin{align} 
\label{eq:tx_default} T_{s} = & T_\text{RTS} + 3\cdot T_\text{SIFS} + T_\text{CTS} + T_\text{DATA} + T_\text{BACK} + T_\text{DIFS} + T_{\mathrm{e}}, 
\end{align}
where
\begin{align} 
T_{\text{DATA}} = T_{\rm PHY} + \left \lceil \frac{L_H+N(L_{D} + L)+ L_{T}}{\text{DBPS}} \right \rceil T_{\rm OFDM},
\end{align}
with $L_H$ as the MAC header, $L_D$ as the MPDU delimiter, $L$ as the MPDU size, and $L_T$ as the tail bits. DBPS denotes the data bits per symbol, which depends on the number of subcarriers (and therefore of the channel width), the number of spatial streams, and the MCS used.

For NPCA transmissions, the maximum transmission duration is determined by the OBSS transmission duration, i.e., $T_{s,\text{OBSS}}$, and includes additional overheads: the time required to switch to the NPCA primary channel ($T_{\text{NPCA}}$) and to return to the original primary channel ($T_{\text{switch}}$). Consequently, the effective transmission time for NPCA is $T_{s,\text{OBSS}} - T_{\text{NPCA}} - T_{\text{switch}}$. 

In both legacy (i.e., NPCA disabled) and NPCA transmissions, the number of packets transmitted is calculated as the maximum number of packets that can be aggregated within the available time, limited by the maximum A-MPDU size ($\Delta$) as described in Section~\ref{sec:model}. 

\subsection{Performance Metrics}

In this paper, we consider two performance metrics as described next.

\subsubsection*{Throughput (bps)} 
\blue{The throughput $\Gamma^{(n)}$ of BSS $n$ is defined as
\begin{equation}
    \Gamma_{n} = (1-\text{PER})\left(\sum_{\forall s \in \Omega : n \in s}{\mu^s_n N^{s}_{n}\pi_s}\right) L,
    \label{eq:throughput}
\end{equation}
where $\mu^s_n N^{s}_{n} L$ is the amount of data bits/second effectively transmitted when the system is in state $s$ by BSS $n$. The PER is applied as a scaling factor.}

\subsubsection*{Channel access delay (ms)} 

To estimate the channel access delay, we perform an event-based simulation based on the $\mathbf{Q}$ matrix.  
Starting from the current state, we identify the earliest upcoming event and transition to the corresponding next state, updating the system time accordingly. Throughout the simulation, we record the time intervals between consecutive transmissions for each BSS. These intervals reflect both the probability of successful channel access and the time spent deferring due to contention, encompassing the total time a BSS spends deferring and transmitting. All simulations begin from state~$0$.

\blue{Note that the channel access delay results obtained from the CTMC model inherently capture the variability of both backoff and transmission durations, which follow exponential distributions. Therefore, in Section~\ref{Sec:Results}, we report only the mean channel access delay values, as they facilitate the interpretation of the results, and provided insights.}

\blue{\begin{table*}[ht!]
    \centering
    \begin{tabular}{c|cc||ccc}
            & \multicolumn{2}{c||}{\textbf{CTMC model}} & \multicolumn{3}{c}{\textbf{Simulation}} \\   
        \hline\hline
        \textbf{Scenario} I & Throughput (Mbps)  & Ch. Access Delay (msecs)  & Throughput (Mbps)   & Ch. Access Delay (msecs) & Coll. Prob\\
        \hline
         BSS A  & 213.9 & 6.05 & 211.6 & 6.09 & 0.1087 \\
         BSS B  & 48.5 & 5.98 & 48.12 & 6.07 & 0.1084 \\
         \hline
        \textbf{Scenario II} & Throughput (Mbps)  & Ch. Access Delay (msecs)  & Throughput (Mbps)  & Ch. Access Delay (msecs) & Coll. Prob\\
        \hline
         BSS A  & 194.9 & 6.65 & 193.3 & 6.67 & 0.110 \\
         BSS B  & 44.1 & 6.55 & 43.8  & 6.66 & 0.109 \\
         BSS D  & 475.0 & 2.70 & 473.5 & 2.72 & 0.000504 \\         
         \hline
        \textbf{Scenario III} & Throughput (Mbps)  & Ch. Access Delay (msecs)  & Throughput (Mbps)  & Ch. Access Delay (msecs) & Coll. Prob\\
        \hline
         BSS A  & 193.6 & 6.68 & 191.9 & 6.72 & 0.111 \\
         BSS B  & 43.8 & 6.72 & 43.5 & 6.72 & 0.110 \\
         BSS C  & 241.9 & 5.39 & 238.9 & 5.40 & 0.112 \\
         BSS D  & 241.9 & 5.41 & 240.4 & 5.37  & 0.111 \\               
         \hline         
    \end{tabular}
    \caption{CTMC model vs simulation results - Legacy operation (Without NPCA)}
    \label{tab:legacy_validation}
\end{table*}}

\blue{\begin{table*}[ht!]
    \centering
    \begin{tabular}{c|cc||ccc}
            & \multicolumn{2}{c||}{\textbf{CTMC model}} & \multicolumn{3}{c}{\textbf{Simulation}} \\   
        \hline\hline
        \textbf{Scenario I} & Throughput (Mbps)  & Ch. Access Delay (msecs)  & Throughput (Mbps)  & Ch. Access Delay (msecs) & Coll. Prob\\
        \hline
         BSS A  & 850.7 & 1.23 & 768.0 & 1.66 & 0.030\\
         BSS B  & 48.5 & 5.99 & 50.22 & 5.72 & 0.104 \\         
         \hline
        \textbf{Scenario II }& Throughput (Mbps)  & Ch. Access Delay (msecs)  & Throughput (Mbps)  & Ch. Access Delay (msecs) & Coll. Prob\\
        \hline
         BSS A  & 375.4 & 2.93 & 369.4 & 3.12 & 0.125 \\
         BSS B  & 44.74 & 6.70 & 45.37 & 6.29 & 0.113 \\
         BSS D  & 360.7 & 3.53 & 338.3 & 3.69 & 0.092 \\     
         \hline
       \textbf{ Scenario III} & Throughput (Mbps)  & Ch. Access Delay (msecs)  & Throughput (Mbps)  & Ch. Access Delay (msecs) & Coll. Prob\\
        \hline
        BSS A  & 277.7  & 4.31  & 268.7 & 3.81 & 0.237 \\
        BSS B  & 39.7 & 7.33 & 39.53 & 6.89 & 0.203 \\
        BSS C  & 245.0 & 4.53 & 228.1 & 4.49  & 0.258 \\
        BSS D  & 212.4 & 6.09 & 210.1 & 5.32 & 0.229\\  
         \hline         
    \end{tabular}
    \caption{CTMC model vs simulation results - NPCA enabled}
    \label{tab:npca_validation}
\end{table*}}

\subsection{Validation} \label{Sec:validation}

\blue{To validate the CTMC model, we compare its throughput and channel access delay results with those obtained from an IEEE 802.11 simulator\footnote{The NPCA simulator is based on the Komondor Wi-Fi simulator: \url{https://github.com/wn-upf/Komondor}}, which has been extended to support NPCA as described in Section~\ref{Sec:Scenarios}. The scenario parameters used for numerical evaluation are as follows: $d_A = 1.5$~m (MCS~11), $d_B = 17$~m (MCS~1), $d_C = 5$~m (MCS~6), and $d_D = 5$~m (MCS~6). The maximum packet aggregation is set to 128 packets. Each simulation run lasts 50 seconds, and the results combine five independent runs using different random number generator seeds, for a total simulation time of 200~s. The simulator implements the standard IEEE~802.11 binary exponential backoff with CW$_\text{min} = 15$ and CW$_\text{max} = 1024$. Transmission durations are deterministic and depend solely on the channel width used and the number of fixed-length packets included.}

\blue{Table~\ref{tab:legacy_validation} presents the results under legacy operation (i.e., when NPCA is disabled). Under these conditions, the CTMC model closely matches the simulation results, with only minor discrepancies. These are primarily attributed to collisions occurring in the simulator, which are not captured by the CTMC model. Collision rates observed in the simulation are consistent with those predicted by Bianchi’s model~\cite{bianchi2000performance}, which estimates a collision probability of approximately 0.11 when two full-buffer APs contend for channel access. It is worth noting that in the simulation, the channel access delay includes the time spent in collisions, as it is measured from the moment a transmission is scheduled until the corresponding Block ACK is received.}

\blue{Table~\ref{tab:npca_validation} presents the results when NPCA is enabled. Overall, both throughput and channel access delay metrics from the model and simulations remain highly accurate. However, some small but noteworthy discrepancies arise. Specifically, BSSs~A and~C (NPCA-capable) achieve slightly lower throughput than expected, while BSSs B and D (non-NPCA-capable) experience slightly higher throughput than anticipated. Although these differences are minor, they reflect an interesting behavior rooted not in collisions only, but in the simulator's implementation of a single backoff instance per BSS, which continues running as the BSS switches between legacy and NPCA modes. In detail, this single-backoff implementation introduces two contrasting effects for NPCA-capable BSSs. First, it reduces their chances of accessing their primary channel. When switching from NPCA back to legacy mode, BSSs~A and~C must draw a new backoff, whereas non-NPCA BSSs~B and~D may resume a paused backoff, gaining an advantage. For instance, in Scenario~I, BSS~B pauses its backoff during BSS~A's transmissions, resulting in a shorter average backoff for subsequent access attempts. Conversely, since BSS~A resets its backoff after NPCA use (happening during BSS~B's transmissions), it is less likely to win next contention against BSS~B on their primary channel, which also limits the number of 160~MHz transmissions it can perform. At the same time, retaining the same backoff instance benefits NPCA access. When BSSs~A and~C switch to the NPCA channel, they continue using the backoff drawn on the primary channel, often resulting in quicker access. In Scenario~II, BSS~D loses more contentions to BSS~A’s NPCA transmissions, slightly reducing its throughput. However, these additional NPCA opportunities only partially offset the loss in legacy transmissions, explaining the observed differences.}

\blue{While the results presented in this section are primarily intended to validate the model's usefulness in analyzing NPCA performance, they also highlight important considerations for the future development of NPCA. In particular, these findings suggest the need for careful design of backoff policies in NPCA operation. Specifically, it is necessary to define how backoff counters should be managed when switching between legacy and NPCA modes. This includes determining whether a single backoff instance should be maintained across both modes, or if separate backoffs should be used for each, along with clear rules for resetting counters and adjusting contention windows in response to collisions. These questions challenge the conventional link between individual transmissions and their associated backoff processes. For instance, if a collision occurs during legacy access but an NPCA opportunity becomes available immediately afterward, should the system double the backoff on the NPCA channel? Or should a new backoff be drawn? Such decisions could significantly impact fairness and efficiency, and thus require careful consideration in future NPCA protocol design.}

\blue{Overall, the results confirm that the CTMC model provides accurate performance estimates---both quantitatively (in terms of throughput and channel access delay) and qualitatively (in capturing the interaction dynamics among contending BSSs).}

\section{Performance Evaluation}
\label{Sec:Results}

In this section, we evaluate the performance of NPCA in the scenarios outlined in Section~\ref{Sec:Scenarios}. We present insights into how NPCA improves Wi-Fi network throughput and reduces channel access delay, and analyze how its activation influences spectrum sharing and interaction among overlapping BSSs~(OBSSs).

\subsection{NPCA Throughput and Channel Access Delay Gains}\label{Sec:npca_gains}

We study the NPCA throughput and delay gains in Scenario~I, where the secondary 80 MHz channel of BSS~A---its NPCA channel---is always available. In this scenario, when NPCA is enabled, BSS~A is able to access its secondary channel (Ch\#2) when its primary channel (Ch\#1) is busy, hence obtaining both higher throughput and lower channel access delay. 

Throughput and channel access delay distributions are derived from multiple scenario instances. In each instance, we randomize two parameters per BSS: $i)$ the station positions, \blue{placing them at random distances uniformly distributed between 1 and 17 meters from their corresponding AP}, with MCSs assigned accordingly, and $ii)$ the maximum A-MPDU value ($\Delta$), ranging from~1 to its upper limit (1024), to introduce high variability in transmission sizes and durations. For reference, in Scenario I, when BSS A uses MCS 11 (1024-QAM, coding rate 5/6) over a 160 MHz channel, it can transmit A-MPDUs of up to 968 packets (each 1400 bytes) within a single TXOP, constrained by the maximum TXOP duration ($T_{\max} = 5$ ms). Similarly, both BSSs~A and B, using MCS~11 on an 80 MHz channel, can transmit up to 484 packets per TXOP.  
 
\begin{figure}[t!]
    \centering
  \begin{subfigure}[b]{0.49\textwidth}
    \includegraphics[width=\linewidth]{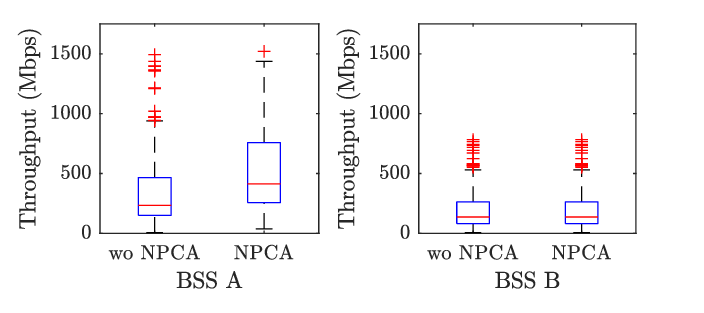}
    \caption{Throughput distribution.} \label{fig:NPCA_F_scen1_S}
  \end{subfigure}\\
  \begin{subfigure}[b]{0.49\textwidth}
    \includegraphics[width=\linewidth]{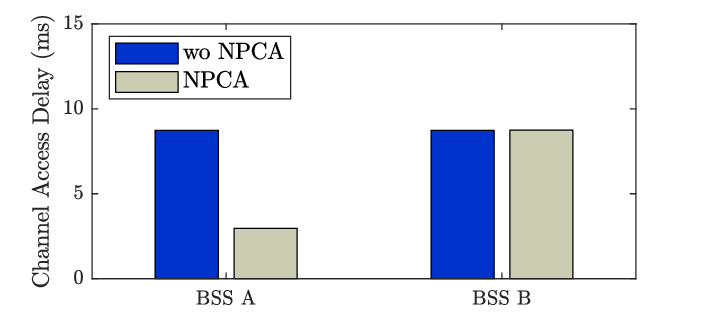}
    \caption{Mean Channel Access Delay.} \label{fig:NPCA_F_scen1_delay}
  \end{subfigure}%
\caption{Throughput and channel access delay results obtained in Scenario I.} \label{fig:NPCA_F_scen1}
\end{figure}

Fig.~\ref{fig:NPCA_F_scen1_S} shows the throughput distribution as a boxplot%
\footnote{\blue{A boxplot shows the median (the line inside the box), the first quartile~(25th percentile) and third quartile~(75th percentile) as the bottom and top edges of the box, respectively. The "whiskers" extend to the most extreme data points within 1.5 times the interquartile range (IQR) from the quartiles, representing the minimum and maximum values within that range. Points outside this range are typically plotted individually as outliers.}}
for BSSs~A and B without and with NPCA enabled. Without NPCA, since BSS A and BSS B share the same primary channel and use the same CW$_{\min}$, following CSMA/CA operation they access the channel the same number of times on average. Therefore, the higher throughput achieved by BSS~A is only because it is able to transmit using the full 160 MHz channel every time it accesses the channel, which turns out on more packets transmitted per TXOP. Enabling NPCA increases the throughput of BSS A by a factor of $\approx \times$1.5, as it can now access its secondary 80 MHz channel (Ch\#2) while BSS B is occupying its primary 80 MHz channel (Ch\#1), whereas before it had to defer. As expected, BSS B is unaffected by BSS A's NPCA transmissions---which occur simultaneously but on a different channel---and therefore, its throughput distribution remains unchanged. \blue{Outliers represent possible but low-probability throughput values, as they depend on specific configuration and scenario parameters for BSSs~A and B.}

Throughput gains come from more frequent channel accesses, and therefore a proportional reduction on the channel access delay should be also expected. \blue{Figure~\ref{fig:NPCA_F_scen1_delay} illustrates the mean delay between two consecutive channel accesses for each BSS. Without NPCA, BSS~A alternates transmissions with BSS~B, accessing the channel approximately every 8.72~ms---comprising 4.36~ms for its own transmission and 4.36~ms for BSS~B's. When NPCA is enabled, BSS~A's mean channel access delay drops to 2.95~ms, corresponding to a reduction factor of approximately 0.338. This improvement occurs because BSS~A can access the channel more frequently: either by winning the contention against BSS~B or by leveraging NPCA transmissions, including performing multiple consecutive NPCA TXOPs each time BSS~B accesses the channel, provided that BSS~B’s transmission duration allows it. This explains why BSS~A’s mean access delay falls below the 4.36~ms value.}
 
\vspace{0.5cm}
\noindent \textbf{Highlight}: NPCA improves throughput and reliability by enabling access to secondary channels when the primary is occupied. Overlapping OBSS transmissions---occupying only the NPCA channel of the NPCA-enabled BSS---remain unaffected, as NPCA operates on different channels, ensuring no negative impact on them. Channel access delay is proportionally reduced as well, enabling faster and more frequent transmissions.  

\subsection{A-MPDU size for Maximum NPCA gain}
 
NPCA transmissions benefit from longer OBSS transmissions. Here, considering Scenario~I, we investigate how different A-MPDU sizes affect NPCA throughput and channel access delay gains. To this end, we evaluate several fixed values of the maximum A-MPDU size, denoted as~$\Delta$. Accordingly, all transmissions now include $\min(M,\Delta)$ packets, where $M$ is the maximum number of packets that can fit in a $T_{\max} = 5$~ms transmission, depending on the employed MCS and channel width. \blue{The distance between each AP and its associated station is uniformly selected at random in each scenario instance between 1~m and 17~m, as in the previous section.}

\begin{figure}[t!]
    \centering
  \begin{subfigure}[b]{0.49\textwidth}
    \includegraphics[width=\linewidth]{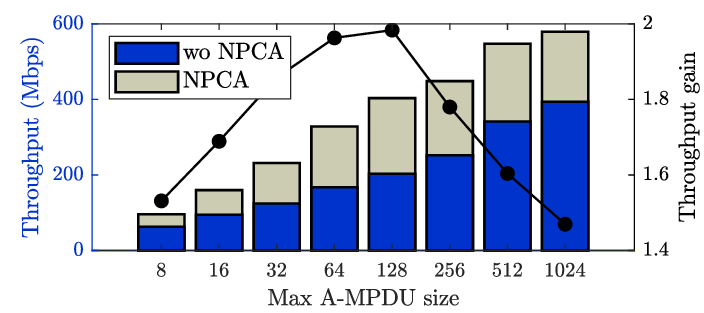}
    \caption{Mean Throughput.} \label{fig:NPCA_AMPDU_S}
  \end{subfigure}\\%
  \begin{subfigure}[b]{0.49\textwidth}
    \includegraphics[width=\linewidth]{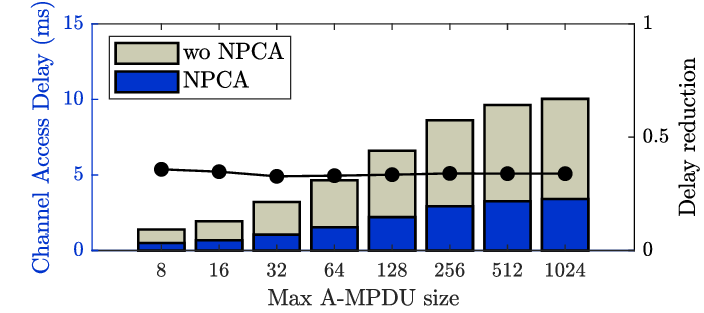}
    \caption{Mean Channel Access Delay.} \label{fig:NPCA_AMPDU_D}
  \end{subfigure}%
\caption{Throughput and channel access delay in BSS A for different maximum A-MPDU ($\Delta$) values in Scenario I.} \label{fig:NPCA_AMPDU}
\end{figure}

Fig.~\ref{fig:NPCA_AMPDU_S} shows the throughput of BSS~A with NPCA disabled~(blue) and enabled~(beige). The \textit{NPCA gain}, defined as the ratio between the throughput with and without NPCA, increases with~$\Delta$. This trend results from longer OBSS transmissions by BSS~B, which create more opportunities for extended NPCA transmissions by BSS~A. The maximum throughput NPCA gain is observed at $\Delta = 128$ packets, where the throughput nearly doubles with NPCA enabled. This is because BSS~A can, on average, transmit as many packets in its NPCA transmissions as it does during legacy 160~MHz transmissions. At lower $\Delta$ values, NPCA overheads limit the throughput gain. At higher $\Delta$ values, the opportunities for NPCA are constrained by BSS~B’s 80~MHz transmissions reaching the maximum TXOP duration (5~ms). In such cases, while BSS~A can include more packets in its 160~MHz transmissions, the throughput of its NPCA transmissions on Ch\#2 (80~MHz) becomes bounded by the same TXOP limit as BSS~B’s. Therefore, the higher throughput observed beyond $\Delta = 128$ packets stems exclusively from the 160~MHz transmissions, since NPCA transmissions are no longer able to scale due to TXOP constraints.

Regarding BSS~A’s mean channel access delay (Fig.~\ref{fig:NPCA_AMPDU_D}), it increases with the maximum A-MPDU size due to longer transmission durations. Interestingly, the ratio between the legacy and NPCA delays remains nearly constant across all A-MPDU sizes, showing the same delay reduction factor discussed earlier in Section~\ref{Sec:npca_gains}.

As for BSS~B, although not shown, its behavior follows the pattern described in the previous section. Increasing the maximum A-MPDU size results in both higher throughput and higher channel access delay. When the A-MPDU size is small enough to allow BSS~B to transmit all packets within the TXOP, its throughput matches that of BSS~A without NPCA. However, once BSS~B can no longer fit as many packets per TXOP, its throughput drops slightly below that of BSS~A without NPCA.

\vspace{0.5cm}
\noindent\textbf{Highlight:} The NPCA throughput gain is bounded by the duration of OBSS transmissions and is thus influenced by the A-MPDU size. Meanwhile, the channel access delay with NPCA remains consistently around one-third of the baseline across all A-MPDU sizes, as NPCA enables near-continuous transmissions, often allowing multiple NPCA transmissions per opportunity when OBSS transmissions are sufficiently long.

\subsection{Overcoming the OBSS Performance Anomaly}

\begin{figure}[t!]
    \centering
  \begin{subfigure}[b]{0.49\textwidth}
    \includegraphics[width=\linewidth]{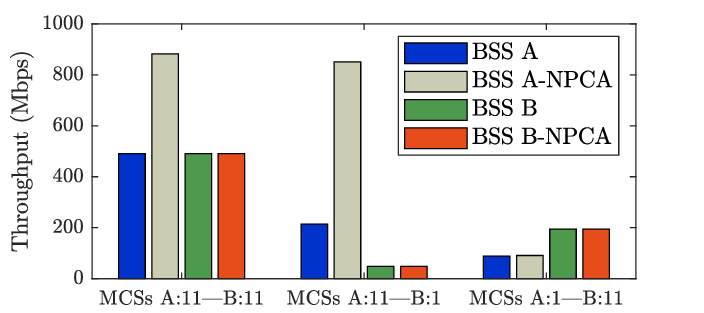}
    \caption{Mean Throughput.} \label{fig:NPCA_OBSS_PA_S}
  \end{subfigure}\\%
  \begin{subfigure}[b]{0.49\textwidth}
    \includegraphics[width=\linewidth]{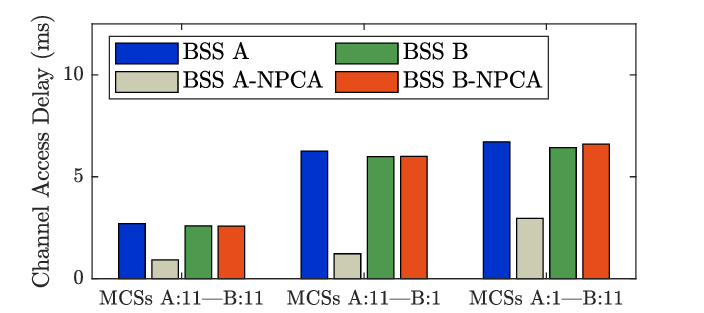}
    \caption{Channel Access Delay.} \label{fig:NPCA_OBSS_PA_D}
  \end{subfigure}%
\caption{Throughput and channel access delay when BSS A and B use different MCSs, without and with NPCA, in Scenario~I.} \label{fig:NPCA_PA}
\end{figure}

In the previous subsections, we analyzed multiple instances of Scenario I, where stations were randomly deployed within the coverage area, resulting in diverse MCSs. Here, we focus on specific cases where BSS A and B use either the lowest MCS (MCS 1, BPSK 1/2) or the highest (MCS 11, 1024-QAM 5/6), with a maximum A-MPDU size of $\Delta=128$ packets. Our primary objective is to assess whether NPCA can mitigate the OBSS \textit{802.11 performance anomaly}---a phenomenon where long OBSS transmissions due to a low MCS degrade the throughput of all OBSSs, including those using higher MCS values. This effect was first described in~\cite{heusse2003performance} for a single, multi-rate BSS.

Fig.~\ref{fig:NPCA_OBSS_PA_S} presents the mean throughput for BSS A and B, with and without NPCA, across three MCS combinations:
\begin{enumerate}
    \item \textit{Both BSSs using MCS 11}: Without NPCA, BSS A and B achieve equal throughput (490 Mbps) as both transmit 128 packets per channel access.  
    With NPCA, BSS~A leverages its secondary 80 MHz channel (i.e., the NPCA primary channel) during BSS B’s transmissions, increasing its throughput to 882 Mbps ($\times$1.8 gain). As expected, BSS B’s throughput remains the same.
    \item \textit{BSS~A using MCS 11, BSS~B using MCS 1}: Without NPCA, BSS~A’s throughput drops from 490 to 213 Mbps because BSS B’s transmission duration increases from 1.58 ms to 5 ms, reducing BSS A’s channel access rate. BSS B, despite its prolonged transmission time, delivers only 29 packets, achieving a low throughput of 48 Mbps. With NPCA, BSS~A exploits BSS B’s extended transmission periods to send more than 128 packets across multiple consecutive NPCA transmissions. Specifically, after completing an NPCA transmission with 128 packets (the A-MPDU limit), BSS~A, recognizing that its primary channel is still occupied by BSS B, initiates additional NPCA transmissions as described in Section~\ref{Sec:NPCA}. This process repeats until the NPCA opportunity ends, boosting BSS A’s throughput to 850 Mbps ($\times$3.9 gain). These results highlight NPCA’s effectiveness in mitigating the negative impact of long OBSS transmissions.
    \item \textit{BSS~A using MCS~1, BSS~B using MCS~11}: Without NPCA, the roles are reversed—BSS~A experiences limited throughput due to its low MCS, which also penalizes BSS~B. Enabling NPCA does not yield any significant gain, as BSS~A is only able to transmit six packets during the short NPCA opportunities.  
\end{enumerate}

Fig.~\ref{fig:NPCA_OBSS_PA_D} shows the mean channel access delay. When one OBSS operates at a high MCS, it reduces the channel access delay for the other, and vice versa. These results are consistent with the throughput analysis, reinforcing NPCA’s effectiveness in addressing the OBSS performance anomaly. It is worth noting that even when no throughput gain is observed---such as when the OBSS performs short transmissions---there is still a clear reduction in channel access delay. This benefits low-latency short transmissions, which can take advantage of the improved channel availability.

\vspace{0.5cm}
\noindent\textbf{Highlight:} NPCA effectively mitigates the OBSS performance anomaly. When one OBSS operates at a low MCS, prolonged transmissions degrade overall network throughput. NPCA enables high-MCS BSSs to exploit these extended NPCA opportunities, significantly improving throughput.

\subsection{Secondary Channel Activity: A Zero-sum Game?}

\begin{figure}[t!]
    \centering
  \begin{subfigure}[b]{0.49\textwidth}
    \includegraphics[width=\linewidth]{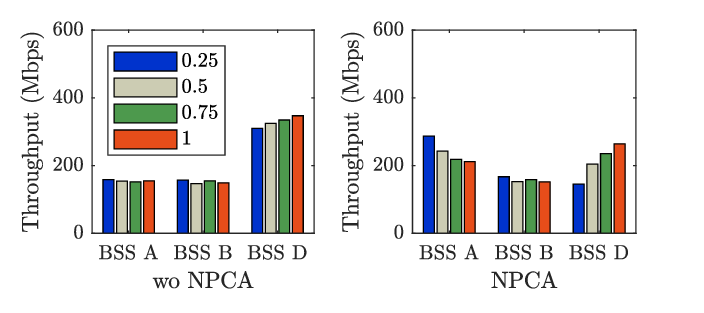}
    \caption{Mean Throughput.} \label{fig:ThreeBSSs}
  \end{subfigure}\\%
  \begin{subfigure}[b]{0.49\textwidth}
    \includegraphics[width=\linewidth]{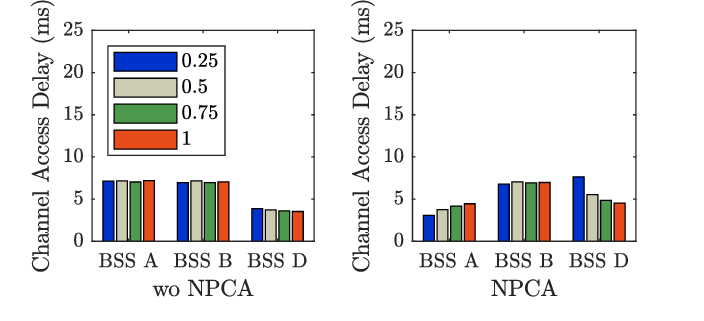}
    \caption{Mean Channel Access Delay.} \label{fig:NPCA_3OBSS_Delay}
  \end{subfigure}%
\caption{Mean throughput and channel access delay for BSSs A, B and D in Scenario~II.} \label{fig:NPCA_OBSSactivity}
\end{figure}

In this section, we investigate the impact of OBSS activity on BSS~A's NPCA channel (Ch\#2) in terms of achievable throughput, considering \textit{Scenario~II}, where BSSs~A, B, and D are active. Specifically, we examine the effect of varying BSS~D's activity levels. To model this, we adjust BSS~D’s contention aggressiveness by scaling its channel access rate~$\lambda$ with the parameter~$\alpha_D$, thereby controlling its channel access intensity (i.e., high (low) values of~$\alpha_D$ correspond to low (high) backoff values). We assume a maximum A-MPDU size of~$\Delta = 128$ packets and \blue{randomize station positions by placing them at distances uniformly distributed between 1 and 17 meters from their corresponding AP.}

In this Scenario II, when NPCA is not enabled, the asynchronous operation among the BSSs significantly limits BSS A’s ability to utilize the full 160 MHz bandwidth. Even at low values of $\alpha_D$, Ch\#2 is almost continuously occupied by BSS~D, forcing BSS~A to transmit only on Ch\#1. In this configuration, BSSs A and B typically alternate access to Ch\#1, while Ch\#2 remains exclusively used by BSS D. However, when NPCA is enabled, BSS A gains additional opportunities to contend with BSS~D. While BSS B occupies Ch\#1, BSS A switches to its NPCA channel (Ch\#2) and attempt to access the medium there, contending with BSS D as mentioned.

Fig.~\ref{fig:ThreeBSSs} (left side) illustrates the mean throughput of BSSs~A, B, and D when NPCA is not enabled. In this scenario, increasing~$\alpha_D$ has an almost negligible effect overall. For BSSs~A and B, increasing~$\alpha_D$ slightly decreases their throughput, as it reduces the likelihood of 160~MHz transmissions by BSS~A. For BSS~D, a higher~$\alpha_D$ reduces its contention time for channel access, thereby also decreasing BSS~A's transmission opportunities on Ch\#2, and resulting in higher throughput for BSS~D. 

The mean throughput of each BSS when NPCA is enabled is shown in Fig.~\ref{fig:ThreeBSSs} (right side). As discussed earlier, NPCA allows BSS~A to directly contend with BSS D for access to Ch\#2 when Ch\#1 is occupied by BSS~B. This ability to compete in both channels increases BSS A's transmission opportunities compared to BSS~B and D, resulting in higher throughput for BSS~A. Furthermore, since BSS~A and BSS~D now compete for the access to Ch\#2, the impact of $\alpha_D$ on BSS A's throughput becomes significant. For low values of $\alpha_D$ (i.e., large backoff intervals for BSS D), BSS A has more chances of accessing Ch\#2, as reflected in the figure. Overall, increasing $\alpha_D$ shifts throughput from BSS A to BSS D. Finally, as expected, BSS B's throughput remains unaffected by enabling NPCA.

When comparing the aggregate throughput---sum of the individual throughput of BSSs A, B and D---with and without NPCA, the difference is minimal. For instance, at $\alpha_D=1$, the aggregate throughput without NPCA is 650 Mbps, which decreases to 626 Mbps with NPCA. With NPCA, two factors contribute to the slight throughput loss: $i)$ increased contention between BSS~A and D on Ch\#2, and $ii)$ the additional transmission overhead introduced by NPCA. Thus, while NPCA does not significantly affect the aggregate throughput, it affects how the throughput is shared among the BSSs.   

Fig.~\ref{fig:NPCA_3OBSS_Delay} presents the mean channel access delay. The delay trends mirror the throughput results, illustrating how enabling NPCA benefits BSS~A at the expense of BSS~D. 

\vspace{0.5cm}
\noindent \textbf{Highlight}: With NPCA enabled, BSSs~A and D contend for access to Ch\#2, making their throughputs and channel access delays mutually sensitive to each other's activity levels. Although NPCA introduces additional contention and overhead, its impact on aggregate throughput remains minimal. Nevertheless, while it represents a zero-sum game in terms of aggregate throughput, NPCA significantly reshapes the utilization of spectrum resources. This underscores the importance of accounting for NPCA in future Wi-Fi channel allocation strategies.

\subsection{Multiple NPCA BSSs Contending}

\begin{figure}[t!]
    \centering
  \begin{subfigure}[b]{0.49\textwidth}
    \includegraphics[width=\linewidth]{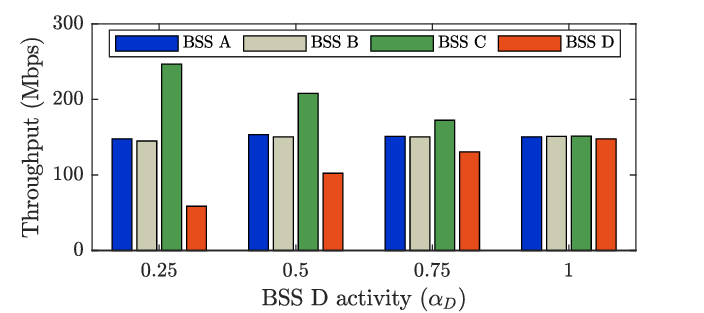}
    \caption{Without NPCA.} 
    \label{fig:Four_BSSs_S}
  \end{subfigure}\\%
  \begin{subfigure}[b]{0.49\textwidth}
    \includegraphics[width=\linewidth]{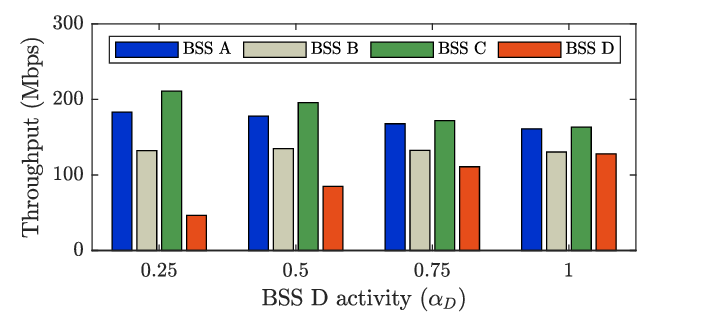}
    \caption{With NPCA.} 
    \label{fig:Four_BSSs_D}
  \end{subfigure}%
\caption{Mean throughput (Mbps) for each BSS without and with NPCA in Scenario~III.} \label{fig:NPCA_OBSSactivity}
\end{figure}

Finally, we consider Scenario III, where BSS~C is activated to analyze a setup in which two BSSs---namely BSS A and BSS C---support NPCA transmissions. In this scenario, BSS~C performs NPCA transmissions on Ch\#1 (its NPCA channel) when its primary channel, Ch\#2, is occupied by BSS~D. As in the previous section, we vary the activity factor of BSS D from $\alpha_D = 0.25$ to $1$. \blue{Stations are uniformly distributed at random between 1 and 17 meters from their respective APs}, and the maximum A-MPDU size is set to $\Delta=128$ packets. Note that we do not present channel access delay results for this scenario, as they do not provide additional insights.

Fig.~\ref{fig:Four_BSSs_D} shows the throughput for each BSS, both without and with NPCA enabled, as the activity factor of BSS D increases. Without NPCA, increasing $\alpha_D$ leads to reduced throughput for BSS C due to increased contention with BSS D on Ch\#2. BSS A and B remain unaffected. At $\alpha_D = 1$, the symmetry of the scenario ensures that all four BSSs achieve the same throughput. With NPCA enabled, BSS~A and BSS~C can now transmit using NPCA on Ch\#2 and Ch\#1, respectively, which makes them to interact. We observe the following: $i)$ for $\alpha_D < 1$, BSS~A benefits from more NPCA transmission opportunities than BSS C because BSS B is more active than BSS D. This results in a significant throughput gain for BSS A, at the expense of BSSs C and D. For instance, at $\alpha_D = 0.25$, BSS A’s throughput increases from 147~Mbps to 183~Mbps, while BSS C’s throughput drops from 246~Mbps to 211~Mbps. As $\alpha_D$ approaches 1, the throughput of BSS~A and C equalizes; $ii)$ BSS B’s throughput decreases with NPCA enabled, since BSS C now competes for Ch\#1. This degradation becomes more pronounced as BSS D’s activity increases, allowing more frequent NPCA transmissions by BSS C on Ch\#1; $iii)$ For $\alpha_D < 1$, as discussed before, BSS C sees reduced throughput under NPCA due to BSS A benefiting more from NPCA opportunities on Ch\#2, driven by BSS B’s higher activity compared to BSS D; and, $iv$) BSS D’s throughput increases with $\alpha_D$ and mirrors the behavior of BSS B when $\alpha_D = 1$.

\vspace{0.5cm}
\noindent \textbf{Highlight:} While OBSS transmissions on a NPCA-enabled BSS’s primary channel may allow it to benefit from additional transmissions on its secondary channel, it is generally preferable---in terms of achievable throughput---to avoid such OBSS activity whenever possible. In symmetric scenarios where avoidance is not feasible, NPCA can still provide a throughput gain, albeit at the expense of non-NPCA BSSs, due to increased contention on their channels.   

\section{Related Work}

NPCA was introduced in the 802.11 Ultra High Reliability~(UHR) Study Group (SG) as a potential feature for Wi-Fi~8, outlining its basic operation for both APs and stations~\cite{NPCA_UHR_2}. Illustrative results (simulation only) to demonstrate potential performance gains were also presented in the UHR SG~\cite{NPCA_UHR_1, NPCA_UHR_3}. Specifically, \cite{NPCA_UHR_1} provided throughput results for scenarios involving two BSSs (1 OBSS) and three BSSs (2 OBSSs), corresponding to our Scenarios I and II. Similarly, \cite{NPCA_UHR_3} reported simulation results for a scenario akin to our Scenario II, considering both full-buffer and finite-load traffic, along with latency measurements. In both works, the results align with the findings in this paper, showing similar improvements in throughput and latency.

The discussion on NPCA continued in the 802.11bn Task Group (TGbn), with a focus on implementation details. Key on-going considerations include the conditions that should trigger a switch when the primary channel is busy~\cite{TGbn_1}, challenges in detecting OBSS transmissions and their bandwidth when not all devices in the NPCA BSS observe them~\cite{TGbn_2}, and the configuration of EDCA parameters (e.g., parameter sets, reuse of backoff counters) and wide-bandwidth transmissions for NPCA~\cite{TGbn_3}.

Outside the 802.11 community, research on NPCA is limited, with the exception of~\cite{wei2024non}, which presents an analytical model of NPCA based on Bianchi’s model~\cite{bianchi2000performance}. Unlike our analysis, the model in~\cite{wei2024non} assumes that all BSSs share the same primary channel (i.e., Scenario I) and does not cover Scenarios II and III. Instead, its focus is on the impact of increasing the number of stations and including uplink transmissions. Nonetheless, the conclusions in~\cite{wei2024non} regarding throughput and latency gains are consistent with ours.

In summary, we advance beyond the state of the art by presenting a comprehensive analysis capable of capturing complex scenarios involving multiple BSSs. Our results account for various critical aspects of Wi-Fi, including the effects of different MCSs and TXOP durations. Furthermore, we highlight the OBSS performance anomaly, demonstrating how NPCA effectively mitigates its negative effects on performance. Additionally, we provide insights into how NPCA can increase contention in high-traffic scenarios, offering a detailed understanding of its trade-offs and potential in future Wi-Fi deployments.

\section{Conclusions}

In this paper, we have studied the Non-Primary Channel Access mechanism, a distinctive feature envisioned for future IEEE~802.11bn Wi-Fi networks. NPCA enables devices to contend and transmit on a secondary channel when the primary channel is occupied by a transmission from an OBSS, thereby reducing channel access delay and improving throughput. Notably, NPCA proves particularly effective in mitigating the OBSS performance anomaly, as low-rate, long-duration transmissions create opportunities for similarly long NPCA transmissions. However, as expected, enabling NPCA also increases contention on secondary channels, potentially degrading performance for BSSs operating on those channels. 

To analyze NPCA operation, we developed a Continuous-Time Markov Chain modeling approach, which offers a valuable framework to characterize the interactions among overlapping BSSs when NPCA is enabled. While remaining tractable, the model provides unique and valuable insights into the potential benefits and limitations of NPCA in dense WLANs. \blue{The validation process has highlighted several implementation challenges related to NPCA, particularly in relation to backoff policies. Our findings demonstrate the value of modeling specific features---even when certain assumptions and simplifications are required---as a means to deepen our understanding of system behavior and the achievable performance limits. Future work may also explore how NPCA overheads can be further mitigated to enhance its efficiency.}

Several open research questions remain. For example, future work could investigate NPCA’s performance under mixed traffic conditions, where NPCA opportunities are only leveraged by low-latency traffic. This approach could leverage NPCA’s reduced channel access delay while minimizing contention with neighboring networks operating on the secondary channel. Moreover, the concept of channel switching could be further generalized by enabling NPCA transmissions to opportunistically utilize any available idle channel, rather than being limited to use the secondary one. This would increase the chances of successful transmissions and further reduce OBSS contention.

\section*{Acknowledgments}

B. Bellalta and F. Wilhelmi were in part supported by Wi-XR PID2021-123995NB-I00 (MCIU/AEI/FEDER,UE), and MdM CEX2021-001195-M (MICIU/AEI/10.13039/501100011033). B. Bellalta is also supported by ICREA Academia 00077.
L. Galati Giordano was in part supported by UNITY-6G project, funded from European Union’s Horizon Europe Smart Networks and Services Joint Undertaking (SNS JU) research and innovation programme under the Grant Agreement No 101192650. 
G. Geraci was in part supported by the Spanish Research Agency through grants PID2021-123999OB-I00, CEX2021-001195-M, and CNS2023-145384. 
%

\bibliographystyle{unsrt}
\bibliography{References}

\begin{thebibliography}{10}

\bibitem{geraci2025wi}
Giovanni Geraci, Francesca Meneghello, Francesc Wilhelmi, David Lopez-Perez, I{\~n}aki Val, Lorenzo~Galati Giordano, Carlos Cordeiro, Monisha Ghosh, Edward Knightly, and Boris Bellalta.
\newblock {Wi-Fi: Twenty-Five Years and Counting}.
\newblock {\em arXiv preprint arXiv:2507.09613}, 2025.

\bibitem{barrachina2021wi}
Sergio Barrachina-Mu{\~n}oz, Boris Bellalta, and Edward~W Knightly.
\newblock {Wi-Fi channel bonding: An all-channel system and experimental study from urban hotspots to a sold-out stadium}.
\newblock {\em IEEE/ACM Transactions on Networking}, 29(5):2101--2114, 2021.

\bibitem{khorov2018tutorial}
Evgeny Khorov, Anton Kiryanov, Andrey Lyakhov, and Giuseppe Bianchi.
\newblock {A tutorial on IEEE 802.11 ax high efficiency WLANs}.
\newblock {\em IEEE Communications Surveys \& Tutorials}, 21(1):197--216, 2018.

\bibitem{galati2024will}
Lorenzo Galati-Giordano, Giovanni Geraci, Marc Carrascosa, and Boris Bellalta.
\newblock {What will Wi-Fi 8 be? A primer on IEEE 802.11 bn ultra high reliability}.
\newblock {\em IEEE Communications Magazine}, 62(8):126--132, 2024.

\bibitem{valwi}
I{\~n}aki Val, David L{\'o}pez-P{\'e}rez, Aleksandra Kijanka, Sigurd Schelstraete, Luis Mu{\~n}oz, Diego Arlandis, and Marcos Mart{\'\i}nez.
\newblock {Wi-Fi} 8 unveiled: Key features, multi-{AP} coordination, and the role of {C-TDMA}.
\newblock {\em techXriv}, 2025.

\bibitem{boorstyn1987throughput}
Robert Boorstyn, Aaron Kershenbaum, Basil Maglaris, and Veli Sahin.
\newblock {Throughput analysis in multihop CSMA packet radio networks}.
\newblock {\em IEEE Transactions on Communications}, 35(3):267--274, 1987.

\bibitem{liew2010back}
Soung~Chang Liew, Cai~Hong Kai, Hang~Ching Leung, and Piu Wong.
\newblock {Back-of-the-envelope computation of throughput distributions in CSMA wireless networks}.
\newblock {\em IEEE Transactions on Mobile Computing}, 9(9):1319--1331, 2010.

\bibitem{laufer2015capacity}
Rafael Laufer and Leonard Kleinrock.
\newblock {The capacity of wireless CSMA/CA networks}.
\newblock {\em IEEE/ACM Transactions on Networking}, 24(3):1518--1532, 2015.

\bibitem{bellalta2015interactions}
Boris Bellalta, Alessandro Checco, Alessandro Zocca, and Jaume Barcelo.
\newblock {On the interactions between multiple overlapping WLANs using channel bonding}.
\newblock {\em IEEE Transactions on Vehicular Technology}, 65(2):796--812, 2015.

\bibitem{faridi2016analysis}
Azadeh Faridi, Boris Bellalta, and Alessandro Checco.
\newblock {Analysis of dynamic channel bonding in dense networks of WLANs}.
\newblock {\em IEEE Transactions on Mobile Computing}, 16(8):2118--2131, 2016.

\bibitem{barrachina2019dynamic}
Sergio Barrachina-Munoz, Francesc Wilhelmi, and Boris Bellalta.
\newblock {Dynamic channel bonding in spatially distributed high-density WLANs}.
\newblock {\em IEEE Transactions on Mobile Computing}, 19(4):821--835, 2019.

\bibitem{wilhelmi2021spatial}
Francesc Wilhelmi, Sergio Barrachina-Mu{\~n}oz, Cristina Cano, Ioannis Selinis, and Boris Bellalta.
\newblock {Spatial reuse in IEEE 802.11 ax WLANs}.
\newblock {\em Computer Communications}, 170:65--83, 2021.

\bibitem{wilhelmi2023throughput}
Francesc Wilhelmi, Lorenzo Galati-Giordano, Giovanni Geraci, Boris Bellalta, Gianluca Fontanesi, and David Nu{\~n}ez.
\newblock {Throughput analysis of IEEE 802.11 bn coordinated spatial reuse}.
\newblock In {\em 2023 IEEE Conference on Standards for Communications and Networking (CSCN)}, pages 401--407. IEEE, 2023.

\bibitem{adame2019tmb}
Toni Adame, Marc Carrascosa, and Boris Bellalta.
\newblock {The TMB path loss model for 5 GHz indoor WiFi scenarios: On the empirical relationship between RSSI, MCS, and spatial streams}.
\newblock In {\em 2019 Wireless Days (WD)}, pages 1--8. IEEE, 2019.

\bibitem{bellalta2016throughput}
Boris Bellalta.
\newblock {Throughput analysis in high density WLANs}.
\newblock {\em IEEE Communications Letters}, 21(3):592--595, 2016.

\bibitem{michaloliakos2016performance}
Antonios Michaloliakos, Ryan Rogalin, Yonglong Zhang, Konstantinos Psounis, and Giuseppe Caire.
\newblock {Performance modeling of next-generation WiFi networks}.
\newblock {\em Computer Networks}, 105:150--165, 2016.

\bibitem{barrachina2019komondor}
Sergio Barrachina-Munoz, Francesc Wilhelmi, Ioannis Selinis, and Boris Bellalta.
\newblock Komondor: A wireless network simulator for next-generation high-density {WLANs}.
\newblock In {\em 2019 Wireless Days (WD)}, pages 1--8. IEEE, 2019.

\bibitem{stojanova2021markov}
Marija Stojanova, Thomas Begin, and Anthony Busson.
\newblock {A Markov model for performance evaluation of channel bonding in IEEE 802.11}.
\newblock {\em Ad Hoc Networks}, 115:102449, 2021.

\bibitem{bianchi2000performance}
Giuseppe Bianchi.
\newblock {Performance analysis of the IEEE 802.11 distributed coordination function}.
\newblock {\em IEEE Journal on Selected Areas in Communications}, 18(3):535--547, 2000.

\bibitem{heusse2003performance}
Martin Heusse, Franck Rousseau, Gilles Berger-Sabbatel, and Andrzej Duda.
\newblock {Performance anomaly of 802.11b}.
\newblock In {\em IEEE INFOCOM}, volume~2, pages 836--843. IEEE, 2003.

\bibitem{NPCA_UHR_2}
{M. Park, et al.}
\newblock {Non-Primary Channel Access}, IEEE 802.11-23/2005r1. 2023.

\bibitem{NPCA_UHR_1}
{Y. Seok, et al.}
\newblock {Non-Primary Channel Access}, IEEE 802.11-23/0797r1. 2023.

\bibitem{NPCA_UHR_3}
{D. Dibakar, et al.}
\newblock {NPC sims follow-up}, IEEE 802.11-23/1444r1. 2023.

\bibitem{TGbn_1}
{V. Ratnam, et al.}
\newblock {Channel Switching Rules for NPCA}, IEEE 802.11-24/1115r1. 2024.

\bibitem{TGbn_2}
{S. Byeon, et al.}
\newblock {NPCA operation issue}, IEEE 802.11-24/1394r0. 2024.

\bibitem{TGbn_3}
{Donju Cha et al.}
\newblock {EDCA for Non Primary Channel Access}, IEEE 802.11-24/0426r0. 2024.

\bibitem{wei2024non}
Dongyu Wei, Liu Cao, Lyutianyang Zhang, Xiangyu Gao, and Hao Yin.
\newblock {Non-primary channel access in IEEE 802.11 UHR: comprehensive analysis and evaluation}.
\newblock In {\em 2024 IEEE 100th Vehicular Technology Conference (VTC2024-Fall)}, pages 1--6. IEEE, 2024.

\end{thebibliography}

\end{document}